\documentclass{epl}
\usepackage{psfig}
\newcommand{\mean}[1]{\left\langle#1\right\rangle}
\newcommand{\dyn}{\mathrm{dyn}}
\newcommand{\eq}{\mathrm{eq}}
\newcommand{\re}{\mathrm{ref}}
\newcommand{\ITEM}{\item}
\title{Shaking a box of sand}
\shorttitle{Shaking a box of sand}
\author{P.F.~Stadler\inst{1}\footnote{E-mail: studla@tbi.univie.ac.at}
\and J.M.~Luck\inst{2}\footnote{E-mail: luck@spht.saclay.cea.fr}
\and Anita Mehta\inst{3}\footnote{E-mail: anita@boson.bose.res.in}}
\institute{
\inst{1}Institut f\"ur Theoretische Chemie und Molekulare Strukturbiologie,
Universit\"at Wien, W{\"a}hringerstra{\ss}e 17, A-1090 Wien, Austria;
and The Santa Fe Institute, 1399 Hyde Park Rd., Santa Fe NM 87501, USA.\\
\inst{2}Service de Physique Th\'eorique\footnote{URA 2306 of CNRS},
CEA Saclay, 91191 Gif-sur-Yvette cedex, France.\\
\inst{3}S N Bose National Centre for Basic Sciences, Block JD Sector 3,
Salt Lake, Calcutta 700098, India;
and ICTP, Strada Costiera 11, 34100 Trieste, Italy.}
\begin{document}
\maketitle
\begin{abstract}
We present a simple model of a vibrated box of sand,
and discuss its dynamics in terms of two parameters
reflecting static and dynamic disorder respectively.
The fluidised, intermediate and frozen (`glassy') dynamical
regimes are extensively probed by analysing the response
of the packing fraction to steady, as well as cyclic,
shaking, and indicators of the onset of glassy behaviour are analysed.
In the `glassy' regime, our model
is exactly solvable, and allows for the qualitative
description of ageing phenomena in terms of two
characteristic lengths; predictions are also made about
the influence of {\em grain shape anisotropy} on ageing behaviour.
\end{abstract}

Vibrating sand results in very varied dynamics, ranging from
glassy~\cite{NowakE:var,Mehta:00} to fluidised \cite{expfl:var,thfl:var}.
Recent experiments, e.g.~\cite{nicolas:00},
have gone some way in validating the
notion~\cite{amgcb:var} that its essential features are captured by models
which incorporate the fast relaxation of individual particles together with
the cooperative rearrangements of clusters.
In this Letter we present a
simple model of a vibrated sand-box, which interpolates between the glassy
and fluidised regimes, and is based on the generalisation of an earlier
cellular automaton (CA) model~\cite{av:var} of an avalanching sandpile.
Our model shows {\em both} fast and slow dynamics in the appropriate regimes:
in particular, it reduces to an exactly solvable model in the frozen
(`glassy') regime, and provides one with a toy model for
ageing in vibrated sand~\cite{jorge:00,letizia:00}.

We consider a rectangular lattice of height $H$ and width $W$ with $N\le HW$
grains located at its lattice points, shaken with vibration intensity $\Gamma$.
Each `grain' is a rectangle with sides $1$ and $a\le 1$, respectively.
Consider a grain $(i,j)$ in row $i$, column $j$ whose height
at any given time is given by $h_{ij} = n_{ij-} + a n_{ij+}$,
with $n_{ij-}$ the number of vertical grains and $n_{ij+}$ the number of
horizontal grains below $(i,j)$:
\begin{itemize}
\ITEM If lattice sites $(i+1,j-1)$, $(i+1,j)$, or $(i+1,j+1)$ are
empty, grain $(i,j)$ moves there with a probability
$\exp(-1/\Gamma)$, in units
such that the acceleration due to gravity, the mass of a grain, and the
height of a lattice cell all equal unity.
\ITEM If the lattice site $(i-1,j)$ below the grain is empty, it will fall
down.
\ITEM If lattice sites $(i-1,j\pm1)$ are empty, the grain at height $h_{ij}$
will fall to either lower neighbour, provided the height difference
$h_{ij}-h_{i-1,j\pm1}\ge 2$.
\ITEM The grain flips from horizontal to vertical with probability
$\exp(-m_{ij}(\Delta H +\Delta h)/\Gamma)$, where $m_{ij}$ is the mass of
the pile (consisting of grains of unit mass) above grain $(i,j)$.
For a rectangular grain, $\Delta H=1-a$ is the height difference
between the initial horizontal and the final vertical state
of the grain.
Similarly, the {\em activation energy} for a flip reads
$\Delta h=b-1$, where $b=\sqrt{1+a^2}$ is the diagonal length of a grain.
\ITEM The grain flips from vertical to horizontal with probability
$\exp(-m_{ij}\,\Delta h/\Gamma)$.
\end{itemize}

\begin{figure}[t]
\centerline{\psfig{file=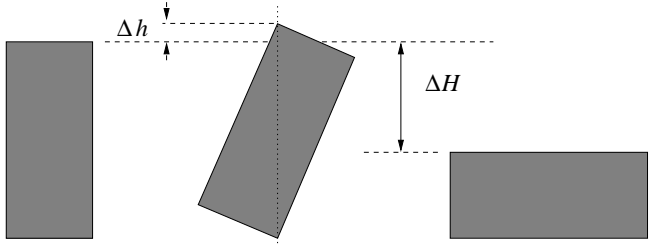,width=0.6\textwidth}}
\caption{A vertical grain needs to be tilted through the height
$\Delta h$ to reach the unstable equilibrium position and flop
to the horizontal, while a horizontal grain needs to be tilted
through an {\it additional} height $\Delta H$ to reach the vertical.
}
\label{fig:fig1}
\end{figure}

Hereafter we consider $\Delta H$ and $\Delta h$ as phenomenological parameters,
which will be seen to be related to the two main characteristic lengths
in the system.

In line with recent investigations of
compaction~\cite{NowakE:var,deGennes:var,sam:var,amgcb:var},
we examine the behaviour
of the packing fraction of our model, as a function of the vibration
intensity $\Gamma$.
Let $N^-$ and $N^+$ be the numbers of vertical and
horizontal grains in the box.
The packing fraction $\phi$ is:
\begin{equation}
\phi=\frac{N^+-aN^-}{N^++aN^-},
\label{phidef}
\end{equation}
which we use as an order parameter reflective of the behaviour of the
compactivity~\cite{sam:var}.
The vertical orientation of a grain thus
wastes space proportional to $1-a$, relative to the horizontal one.

\begin{figure}[t]
\centerline{\psfig{file=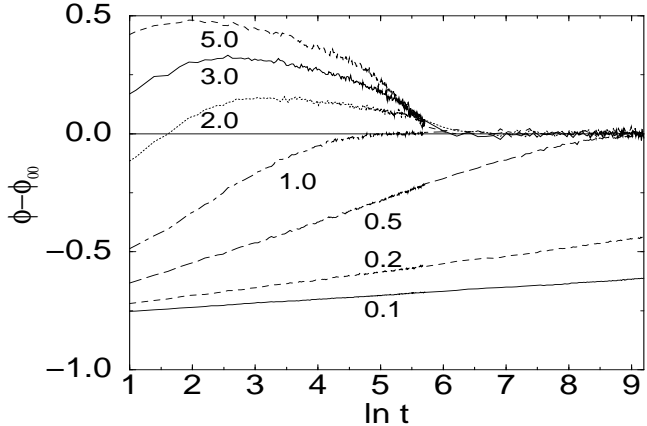,width=0.6\textwidth}}
\caption{Plot of $\phi-\phi_{\infty}$ versus $\ln t$, for different values
of $\Gamma$, indicated on the curves.
Note that
$\phi_{\infty}$ decreases with increasing $\Gamma$, and is thus {\em
distinct} for each curve.}
\label{fig:fig2}
\end{figure}

We examine the response of the packing fraction for $\Delta H = 0.3$,
$\Delta h= 0.05$ to shaking at varying intensities in Fig.~\ref{fig:fig2}.
Since the `equilibrium' packing fraction $\phi_{\infty}$ (which we
determined in separate runs) decreases with increasing
intensity~\cite{amgcb:var}, we plotted the difference $\phi-\phi_{\infty}$ as a
(logarithmic) function of time $T$ in the figure, starting with the {\it
same} initial packing fraction in each case.

The dynamical response of the shaken sand-box includes three distinct
regions, each illustrated by representative curves in the figure:
\begin{itemize}
\ITEM a fluidised region (for $\Gamma\gg 1$), where we observe an initial
increase (caused by a {\it non-equilibrium} and transient `ordering' of
grains in the boundary layer) of the packing fraction that quickly relaxes
to the equilibrium values $\phi_{\infty}$ in each case.
This over-shooting
effect in Fig.~\ref{fig:fig2} increases with $\Gamma$, since grains ever
deeper in the sand-box can now overcome their activation energy to relax to
the horizontal.
This {\it inhomogeneous relaxation} has been seen in
earlier, off-lattice simulations of compaction~\cite{unpublished:00}.
Analogous effects have also been observed in Ref.~\cite{johannes:00}.
\ITEM an intermediate region (for $\Gamma\approx 1$), where the packing
fraction remains approximately constant in the bulk, while the surface
equilibrates via the fast dynamics of {\em single-particle relaxation}.
The specific $\phi_{\infty}$ at which this occurs (0.917 here), is the {\em
single-particle relaxation threshold density} observed in
Ref.~\cite{johannes:00}; non-equilibrium, non-ergodic, fast dynamics allows
single particles locally to find their equilibrium configurations at this
density.
Analogous effects have been observed in recent experiments on
colloids~\cite{science:00}, where the correlated dynamics of {\em fast}
particles was seen to be responsible for most relaxational behaviour before
the onset of the glass transition.
\ITEM a frozen region (for $\Gamma\ll 1$), where the slow dynamics of the
system results in a {\em logarithmic growth} of packing fraction
with time:
\begin{equation}
\phi-\phi_{\infty} = b(\Gamma)\ln t + a,
\end{equation}
where $b(\Gamma)$ increases with $\Gamma$, in good agreement with
experiment~\cite{NowakE:var}.
The slow dynamics has been identified~\cite{johannes:00} with
a cascade process, where the free volume released by the relaxation of one
or more grains allows for the ongoing relaxation of other grains in an
extended neighbourhood.
It includes the phenomenon of {\it bridge collapse},
which, for low vibration intensities, has been seen to be a major mechanism
of compaction~\cite{amgcb:var}.
As $\Gamma$ decreases, the corresponding
$\phi_{\infty}$ increases asymptotically towards the jamming limit
$\phi_{\mathrm{jam}}$, identified with a {\it dynamical phase transition} in
related work~\cite{johannes:00}.
\end{itemize}

We next investigate the analogue of `annealed cooling', where $\Gamma$ is
increased and decreased cyclically, and the response of the packing
fraction observed~\cite{NowakE:var}.
The results obtained
here are similar to those~\cite{unpublished:00} seen using more realistic
models of shaken spheres, but the simplicity of the present model allows
for greater transparency.

\begin{figure}[t]
\centerline{\psfig{file=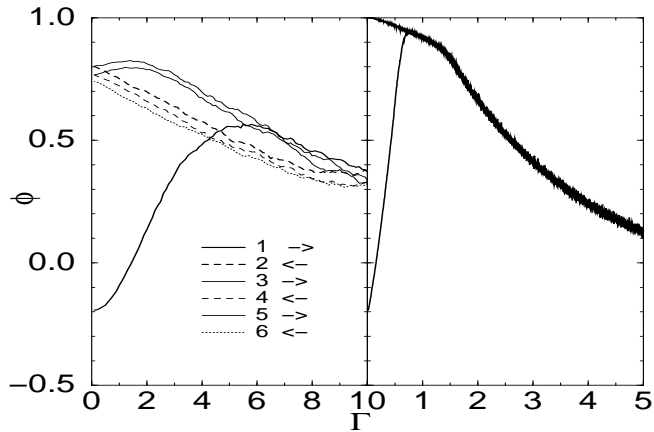,width=0.6\textwidth}}
\caption{Hysteresis curves.
Left: $\Delta\Gamma=0.1$,
$t_{\mathrm{tap}}= 2000$ time units.
Right: $\Delta\Gamma=0.001$,
$t_{\mathrm{tap}}= 10^5$ time units.
Note the approach of the irreversibility point $\Gamma^{*}$ to the `shoulder'
$\Gamma_{\mathrm{jam}}$, as the ramp rate
$\delta\Gamma/t_{\mathrm{tap}}$ is lowered.}
\label{fig:fig3}
\end{figure}

Starting with the sand in a fluidised state, as in the
experiment~\cite{NowakE:var}, we submit the sand-box to taps at a given
intensity $\Gamma$ for a time $t_{\mathrm{tap}}$ and increase the
intensity in steps of $\delta\Gamma$; at a certain point, the cycle is
reversed, to go from higher to lower intensities.
The entire process is then iterated twice.
Figure~\ref{fig:fig3} shows the resulting behaviour of
the volume fraction $\phi$ as a function of $\Gamma$, where an
`irreversible' branch and a `reversible' branch of the compaction curve are
seen, which meet at the `irreversibility point' $\Gamma^{*}$~\cite{NowakE:var}.
The left- and right-hand side of Figure~\ref{fig:fig3} correspond
respectively to high and low values of the `ramp rate'
$\delta\Gamma/t_{\mathrm{tap}}$~\cite{NowakE:var}.
As the ramp rate is lowered, we note that:
\begin{itemize}
\ITEM the width of the hysteresis loop in the so-called reversible branch
decreases.
The `reversible' branch is thus not reversible at all; more
realistic simulations of shaken spheres~\cite{amgcb:var} confirm
the first-order,
irreversible nature of the transition, which allows the density to attain
values that are substantially higher than random close packing, and quite
close to the crystalline limit~\cite{Mehta:00}.
Precisely such a transition
has also recently been observed experimentally in the compaction
of rods~\cite{villaruel:00}.
\ITEM the `irreversibility point' $\Gamma^{*}$ approaches
$\Gamma_{\mathrm{jam}}$ (the shaking intensity at which the jamming limit
$\phi_{\mathrm{jam}}$ is approached), in agreement with results on other
discrete models~\cite{coniglio:00}.
\end{itemize}

\begin{figure}[t]
\centerline{\psfig{file=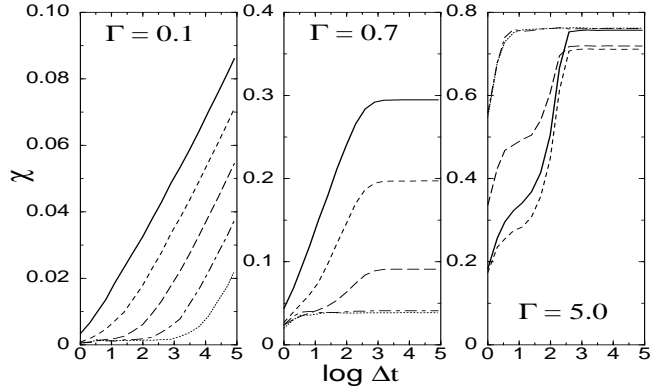,width=0.6\textwidth}}
\caption{Overlap functions $\chi(t_\re,\Delta t)$,
equation~(\ref{eq:chi}), for $\Gamma=0.1$, $0.7$, $5$.
Line styles distinguish five reference times from
$t_\re=1$ (full line) to $t_\re=10^4$ (dotted line).
The time unit is defined as $HW$ attempted Monte Carlo moves.}
\label{fig:fig4}
\end{figure}

The simplicity of our model also permits us to explore the onset of 
`glassy' behaviour between the regimes where fast and slow dynamics
respectively predominate.
We explore this via a configurational overlap function
\begin{equation}
\label{eq:chi}
\chi(t_\re,\Delta t)
=\frac{1}{N}\sum_{i,j}
\Theta[B_{i,j}(t_\re),B_{i,j}(t_\re+\Delta t)].
\end{equation}
Here $B_{i,j}(t)$ can take three distinct values
depending on whether the lattice site ($i,j$) at
time~$t$ is (a) empty, (b) occupied by a $+$ grain,
(c) occupied by a $-$ grain.
We write $\Theta[X,Y]=1-\delta_{X,Y}$; i.e., $\Theta[X,Y]=0$ if
$X=Y$.
$\Delta t$ is the time lag.
Figure~\ref{fig:fig4} shows results for
different values of $\Gamma$, for $\Gamma = 0.1, 0.7, 5$.

The left-hand panel ($\Gamma = 0.1$) shows the logarithmic behaviour
characteristic of ageing, while the right-hand panel ($\Gamma = 5$) shows
the quick equilibration virtually independent of waiting times, which
characterises the fluidised regime.
The middle panel ($\Gamma = 0.7$)
exemplifies the behaviour characteristic of the transition between
the two regimes: there is an
apparent `equilibration' into different metastable states depending on the
waiting time $t_\re$.

The frozen regime is characterised by an absence of holes within the
sand-box, and negligible surface roughness.
Here, our model reduces to an
exactly solvable model of $W$ independent columns of $H$ noninteracting
`grains' $\sigma_n(t)=\pm1$, with $\sigma=+1$ denoting a
horizontal grain, and $\sigma=-1$ denoting a vertical grain.
The orientation of the grain at depth $n$, measured from the top of the system,
evolves according to a Markov dynamics with depth-dependent rates
\begin{equation}
\left\{\matrix{
w(-1\to+1)=\exp(-n\Delta h/\Gamma),\hfill\cr
w(+1\to-1)=\exp(-n(\Delta H+\Delta h)/\Gamma),\hfill}\right.
\end{equation}
as $m_{ij}=n=H+1-i$.

The order parameter describing the mean orientation, which
we hereafter refer to as `orientedness',
$\overline{M}(t)=(1/H)\sum_{n=1}^H M_n(t)$, with $M_n(t)=\mean{\sigma_n(t)}$,
is related to the packing fraction of equation~(\ref{phidef}) as
\begin{equation}
\overline{M}=\frac{(1+a)\phi-(1-a)}{1+a-(1-a)\phi}.
\end{equation}

At equilibrium, the orientedness profile is given by
$M_{n,\eq}=\tanh\big(n/(2\xi_\eq)\big)$, while the local equilibration time
diverges exponentially with depth $n$, as
$\tau_{n,\eq}\approx\exp(n/\xi_\dyn)$.
These expressions involve two
characteristic lengths of the model, the equilibrium length $\xi_\eq$ and
the dynamical length $\xi_\dyn$, which read
\begin{equation}
\xi_\eq=\frac{\Gamma}{\Delta H},\quad\xi_\dyn=\frac{\Gamma}{\Delta h}.
\end{equation}

In the scaling regime where the height $H$ and both lengths $\xi_\eq$,
$\xi_\dyn$ are large, the mean orientedness is
$\overline{M}_\eq\approx(2\xi_\eq/H)$ $\ln\cosh(H/(2\xi_\eq))$.
For $H\ll\xi_\eq$, $\overline{M}_\eq\approx H/(4\xi_\eq)\ll1$:
the system is very weakly ordered, even at equilibrium.
For $H\gg\xi_\eq$, $\overline{M}_\eq\approx1-(2\ln2)\xi_\eq/H$:
the system is strongly ordered at
equilibrium, except for its top skin layer, whose depth is of order $\xi_\eq$.

As the equilibration time diverges exponentially with the depth,
orientational order propagates down the system logarithmically slowly.
More specifically, for
a large but finite time $t$, only a top layer up to an `ordering length'
$\Lambda(t)$ has equilibrated, with
\begin{equation}
\Lambda(t)\approx\xi_\dyn\ln t.
\end{equation}
We have $M_n(t)\approx M_{n,\eq}$ for $n\ll\Lambda(t)$, whereas
$M_n(t)\approx0$ for $n\gg\Lambda(t)$.
The most ordered grains are situated
at a depth comparable to $\Lambda(t)$; they have a maximum orientedness
$M_{\max}(t)\approx\tanh\big((\omega/2)\ln t\big)$, where
\begin{equation}
\omega=\frac{\xi_\dyn}{\xi_\eq}=\frac{\Delta H}{\Delta h}=\frac{1-a}{b-1}
\end{equation}
is the ratio of both characteristic lengths.

The length $\xi_\eq$ is the length upto which disorder persists
in the granular material {\em when it has attained equilibrium}.
The length $\xi_\dyn$ determines the length {\em to which order has propagated}
in the granular material in the glassy regime.
For grains where there are nearly equivalent orientations,
even at equilibrium, there will be a large number of
`disordered' configurations in the top layer, since
these will be almost equivalent to the strictly ordered one.
(An extreme
example would be $a\to1$, which approximately corresponds to the packing
of spheres, which are known to be disordered
even in the nominally `equilibrium' state of random close packing.)
The length $\xi_\dyn$ controls the rate at which order propagates
as a function of time in the glassy regime of a compacted powder.
Both lengths, in experimental terms, thus have the interpretation
of {\it the depth of the boundary layer}
in a vibrated granular system; in the first
case, this description applies when equilibrium has been reached,
while in the second case, this applies to the nonequilibrium
evolution of a vibrated granular bed.

In order for the model to exhibit interesting non-equilibrium or ageing
effects, one must have $\Lambda(t)\ll H$.
The two-time quantities we
investigate to explore ageing are the full two-time correlation function,
$S_n(t,s)=\mean{\sigma_n(t)\sigma_n(s)}$, and the connected one,
$C_n(t,s)=S_n(t,s)-M_n(t)M_n(s)$, with $0\le s$ (waiting time) $\le t$
(observation time).
In terms of the overlap function of equation~(\ref{eq:chi}),
we have $\overline{S}(t_\re+\Delta t,t_\re)=1-2\chi(t_\re,\Delta t)$.
We are led to consider two different non-equ\-i\-li\-brium regimes.
In each case, the mean observables can be expressed, after some algebra,
in terms of the ordering lengths only (see Table~\ref{tab:regimes}).
\begin{table}[t]
\caption{Two different non-equ\-i\-li\-brium regimes.}
\label{tab:regimes}
\begin{center}
\begin{tabular}{l l l}
\hline
& {\bf Regime I} & {\bf Regime II}\\
\hline
& $\xi_\eq\ll\Lambda(t)\ll H\!$ & $\Lambda(t)\ll\xi_\eq,\,H$\\
&\quad $(\omega\ln t\gg1)$ &\quad $(\omega\ln t\ll1)$\\
\hline\noalign{\vskip 1mm}
$\overline{M}(t)$
&$\Lambda(t)/H$
&$\left[\Lambda(2t)\right]^2/(4H\xi_\eq)$\\
$\overline{S}(t,s)$
&$1-[\Lambda(t)-\Lambda(s)]/H$
&$1-\left[\Lambda(2(t-s))\right]/H$\\
$\overline{C}(t,s)\quad$
&$1-[2\Lambda(t)-\Lambda(t+s)]/H$
&$1-\left[\Lambda(2(t-s))\right]/H$\\
\hline
\end{tabular}
\end{center}
\end{table}
In Regime~I, the maximal ordering is very close to perfect, as
$1-M_{\max}(t)\sim t^{-\omega}\ll1$.
This is the conventional frozen regime
(to which our data in Figure~\ref{fig:fig4} correspond).
The top layer of
the system is strongly ordered, most of the grains are flat, and likely to
stay that way: the ageing phenomenon corresponds to the slow ordering
attempts of grains deeper in the bulk, quantified by the logarithmic growth
of the ordering length $\Lambda(t)$.
Table~\ref{tab:regimes} shows that
the mean orientedness is nothing but the fraction $\Lambda(t)/H$ of the
system that has equilibrated.
The two-time correlations are non-stationary,
and they involve $\Lambda(s)$, $\Lambda(t)$, and $\Lambda(t+s)$.

In Regime~II, the maximal ordering is very weak, as
$M_{\max}(t)\approx(\omega/2)\ln t\ll1$.
This regime exists only for
$\omega\ll1$, i.e., $a\to1$ in the geometrical model.
It corresponds to an
even slower dynamics, since now any attempts at ordering are hindered
additionally by a strong probability that a horizontal grain will flip to
the vertical orientation.
Table~\ref{tab:regimes} shows
that the mean orientedness involves the square of the ordering length,
while the two-time correlations do not exhibit {\em any}
non-stationary features characteristic of ageing, at least to leading order,
in this scaling regime.

The physical difference between the two scenarios is comprehensible in
terms of disorder in grain shapes.
Where grains are very irregularly
shaped (Regime~I), the non-equilibrium regime will carry all the usual
characteristics of ageing.
Where, however, grains are regularly shaped
(Regime~II), the signatures of ageing will be hard to detect even in a
highly non-equilibrium regime.
It would be interesting to test this
experimentally: would ageing experiments carried out separately on
weakly vibrated rods (Regime~I) or spheres (Regime~II) have different results?

In conclusion, the simplicity of our model makes it a useful conceptual
tool for probing the dynamical responses of vibrated
sand, from the fluidised to the frozen regimes.
In the latter case, our model is exactly solvable, which allows one to describe
the by now well-established picture of logarithmic compaction,
in terms of two characteristic lengths.
The improvement of this necessarily qualitative picture of ageing
by the addition of more realistic and complex interactions, while
still retaining the overall conceptual simplicity
of our model, constitutes the focus of current research.

\end{document}